\documentclass[showpacs,amsmath,amssymb,floatfix,prl,twocolumn]{revtex4}
\usepackage{graphicx,amsmath}

\begin{document}

%\title {Stabilization of ballistic edge channels by microwave radiation}
%\title {Edge transport mechanism for zero resistance states}
\title {Microwave stabilization of edge transport and zero-resistance states}

\author{ A.D.~Chepelianskii$^{(a)}$ and D.L.Shepelyansky$^{(b,c)}$}
\affiliation{$(a)$ LPS, Universit\'e Paris-Sud, CNRS, 
UMR 8502, F-91405, Orsay, France }
\affiliation{$(b)$ Universit\'e de Toulouse, UPS,
Laboratoire de Physique Th\'eorique (IRSAMC), F-31062 Toulouse, France}
\affiliation{$(c)$ CNRS, LPT (IRSAMC), F-31062 Toulouse, France}

\date{May  5, 2009}

% 73.40.-c Electronic transport in interface structures 
% 05.45.-a Nonlinear dynamics and chaos
% 72.20.My Galvanomagnetic and other magnetotransport effects 

\begin{abstract}
Edge channels play a crucial role for electron transport in 
two dimensional electron gas under magnetic field. 
It is usually thought that ballistic transport 
along edges occurs only in the quantum regime with 
low filling factors.
We show that a microwave field can stabilize edge trajectories even in the 
semiclassical regime leading to a vanishing longitudinal resistance. 
This mechanism gives a clear physical interpretation 
for observed zero-resistance states.
\end{abstract}

\maketitle
The experimental observation of microwave induced zero-resistance 
states (ZRS) in high mobility two dimensional electron gas (2DEG) 
\cite{mani2002,zudov2003} 
attracted significant experimental and theoretical interest. 
Several theoretical explanations have been proposed so far, 
which rely on scattering mechanisms inside the bulk of 2DEG. 
The ``displacement'' mechanism originates from the effect 
of microwaves on disorder elastic scattering in the sample 
\cite{ryzhii,girvin,vavilov,platero}, 
while the ``inelastic'' mechanism involves inelastic processes 
that lead to a modified out-of equilibrium distribution function 
\cite{polyakov,mirlin}. 
Even if these theories reproduce certain experimental features 
we believe that the physical origin of ZRS is still not captured. 
Indeed several arguments can challenge these approaches. 
The above theories naturally generate negative resistance states 
but one has to rely on an uncontrolled out of equilibrium 
compensation of all currents 
to produce ZRS as observed in experiments \cite{mani2002,zudov2003}. 
Also ZRS is observed in very clean samples, therefore in the bulk 
an electron moves like an oscillator 
where selection rules allow transitions only between nearby oscillator states.
Hence resonant transitions are possible only at cyclotron resonance
where the ratio $j$ between microwave frequency $\omega$
 and cyclotron frequency $\omega_c$
is unity. However experiments show that the onset of ZRS occurs
 also for high $j = \omega / \omega_c$
approximately at $j = 1+1/4,2+1/4,...$. High $j$ resonances could 
appear due to nonlinear effects, however the microwave fields are relatively 
weak giving a ratio $\epsilon$ between oscillating component 
of electron velocity 
and Fermi velocity $v_F$ of the order of few percents. 
Therefore the appearance of high $j$ ZRS in ``displacement'' models with weak 
disorder seems problematic. 
In the ``inelastic'' models one assumes that 2DEG evolves in a far from
equilibrium state due to small energy relaxation rates. 
However since the microwave frequency is high compared to the elastic rate 
2DEG has mainly imaginary high frequency conductivity 
and should not significantly 
absorb microwave power. This can be seen very clearly in \cite{mani2002} 
where the amplitude of the Shubnikov-de Haas oscillations is not changed 
by the presence of microwave radiation even when power is 
strong enough to generate ZRS. Hence it seems unlikely that 2DEG 
actually reaches the out of equilibrium states needed for 
the ``inelastic'' theories.

In order to develop a theory for ZRS we note that they occur when 
the mean free path $l_e$ is much larger than the cyclotron 
radius $r_c = v_F / \omega_c$.
In usual 2DEG samples with lower mobilities this regime corresponds to 
strong magnetic fields and quantum Hall effect. In this case it is known 
that propagation  along sample edges is ballistic and 
plays a crucial role in magnetotransport.
It leads to quantization of the Hall resistance $R_{xy}$ 
and to the disappearance of four terminal resistance $R_{xx}$ 
strikingly similar to ZRS \cite{halperin}. 
This occurs at low filling factors $\nu$ when a gap forms 
in the 2DEG density of states due to discreetness of Landau levels.  
In contrast to that ZRS appear at $\nu \simeq 50$ where 
Landau levels are smeared out by disorder. 
Even in this semiclassical regime, edge trajectories 
are still important for transport. 
Guiding along sample edges can lead to a significant decrease of $R_{xx}$
with magnetic fields giving a negative magnetoresistance and 
singularities in $R_{xy}$ \cite{roukes,beenakker}
(note that negative magnetoresistance 
is also observed in ZRS samples \cite{mani2002,zudov2003,bykov}).
This behaviour can be understood theoretically 
from the transmission probability $T$ between  
voltage probes in a Hall bar geometry \cite{beenakker}.
The drop in $R_{xx}$ is linked to increased $T$,
but  transmission remains smaller than unity due to 
disorder and $R_{xx}$ remains finite. 
Recently this model was extended to understand experimental deviations from 
Onsager reciprocity relations in samples under microwave driving \cite{lps}. 
But the impact of microwaves on stability of edge channels was never 
considered before. 

In this Letter we show that microwave radiation can stabilize guiding 
along sample edges leading to a ballistic transport regime 
with vanishing $R_{xx}$
and transmission exponentially close to unity. It was
established experimentally that edge channels are 
very sensitive to irradiation \cite{merz}
and recent contact-less measurements in the ZRS regime 
did not show a significant drop of $R_{xx}$ \cite{kukushkin}
that supports our edge transport mechanism for ZRS. 
Our model also relies on the fact that scattering occur 
on small angles in 2DEG \cite{mani2002,ando}.
This contrasts with other ZRS models which do not rely 
on specific physical properties of 2DEG.

\begin{figure}
\begin{center}
\includegraphics[clip=true,width=8cm]{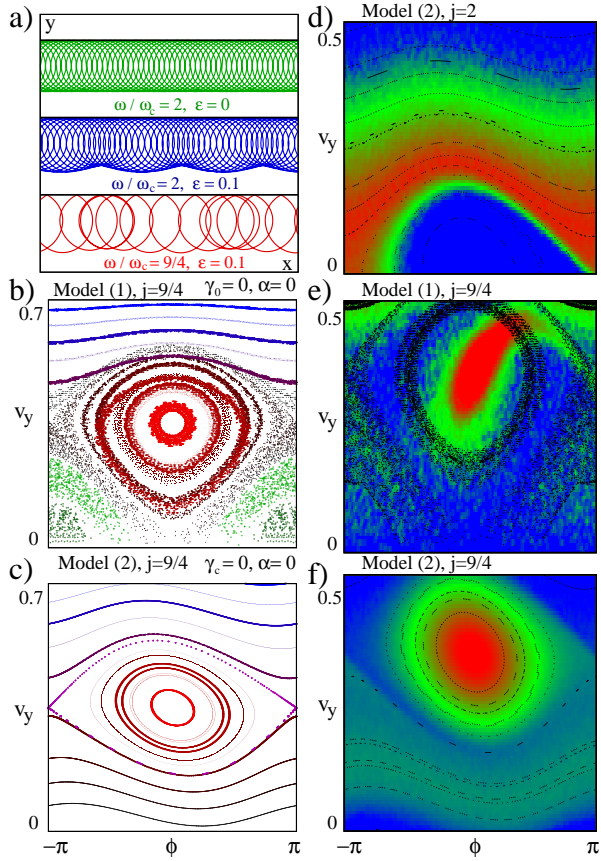}
\vglue -0.3cm
%\leavevmode
%\epsfxsize=8cm
\caption{ (Color online) a) Examples of electron trajectories along sample edge 
for several values of $j=\omega/\omega_c$ and $y$-polarized field $\epsilon$.
b) Poincar\'e section of (\ref{newton}) for $\omega/\omega_c = 9/4$ 
at $y$-polarized $\epsilon = 0.02$.
c) Poincar\'e section in the same region for 
the Chirikov standard map (\ref{chirikovmap}) giving approximate 
description of dynamics in b).
In a,b,c) dissipation and impurity scattering angle are zero. 
d,e,f) Density of propagating particles on the Poincar\'e section 
in presence of noise and dissipation
(red/gray for maximum and blue/black for zero), 
black points show trajectories without noise and dissipation. 
For $\omega/\omega_c = 2$ microwave repels particles from the edge (d),
while for $\omega/\omega_c = 9/4$ particles are trapped inside 
the nonlinear resonance (e,f).
Here $\gamma_0 = 10^{-3}$ (e), $\gamma_c = 10^{-2}$ 
(d,f) and $\alpha \simeq 5 \times 10^{-3}$. 
\label{fig1}}
\end{center}
\end{figure} 

Since filling factors are large we study classical 
dynamics of an electron at the 
Fermi surface \cite{beenakker} propagating along 
a sample edge modeled as a specular wall. 
The motion is describe by Newton equations:
\begin{align}
d \mathbf{v}/d t = \mathbf{\omega_c} \times \mathbf{v} + 
\mathbf{\epsilon} \cos \omega t -
\gamma(v) \mathbf{v} + I_{wall} + I_{S}
\label{newton}
\end{align}
where $\mathbf{\epsilon} = e \mathbf{E}/ (m \omega v_F)$ 
describes microwave driving
field $\mathbf{E}$, velocity is measured in units of Fermi velocity $v_F$,
and $\gamma(v) = \gamma_0 (|\mathbf{v}|^2 - 1)$ describes relaxation 
processes to the Fermi surface. 
The last two terms account for 
elastic collisions with the wall and small angle scattering.
Disorder scattering is modeled as random rotations of $\mathbf{v}$ by 
small angles in the interval $\pm \alpha$ with Poissonian distribution 
over microwave period. 
Examples of electron dynamics along the sample edge 
for $\gamma_0 = 0$ and $\alpha = 0$
are shown in Fig.~1a. They show that even a weak field 
$\epsilon = 0.1$ has strong 
impact on dynamics along the edge. 
A more direct understanding of the dynamics can 
be obtained from the Poincar\'e sections
constructed for the microwave field phase
$\phi = \omega t ({\rm mod} 2 \pi)$ and 
the velocity component $v_y > 0$ at the moment of collision with the wall. 
The system (\ref{newton}) has two and half degrees of 
freedom and therefore the curves on 
the section are only approximately invariant (Fig.~1b). 
The main feature of this
figure is the appearance of a nonlinear resonance. 
We assume for simplicity that 2DEG is not at cyclotron 
resonance and polarization is mainly along $y$ axis.
Since Eq.~(\ref{newton}) is linear outside the wall, 
one can go to the oscillating frame where electron moves on a circular orbit 
while the wall oscillates in $y$ with velocity $\epsilon \sin \omega t$. 
Hence collisions change $v_y$ by twice the wall velocity. 
For small collision angles the time between collisions is 
$\Delta t = 2 (\pi - v_y) / \omega_c$. This yields 
an approximate dynamics description 
in terms of the Chirikov standard map \cite{chirikov}:
\begin{align}
{\bar v_y} = v_y + 2 \epsilon \sin \phi + I_{cc}, \;
{\bar \phi} = \phi + 2 (\pi - {\bar v_y}) \omega_ /\omega_c 
\label{chirikovmap}
\end{align}
The term $I_{cc} = - \gamma_c v_y + \alpha_n$ describes 
dissipation and noise, bars denote values after map iteration
 ($-\alpha < \alpha_n <\alpha$). Damping from electron-phonon and
electron-electron collisions contribute to $\gamma_c$.
The Poincar\'e sections for Eqs.~(\ref{newton},\ref{chirikovmap}) 
are compared in Figs.~1b,c showing that the Chirikov standard map 
gives a good description 
for edge  dynamics under microwave driving. 
A phase shift by $2 \pi$ does not change 
the behavior of map (\ref{chirikovmap}) and hence 
the phase space structure 
is periodic in $j = \omega/\omega_c$ with period unity which naturally 
yields high harmonics. The resonance is centered 
at $v_y = \pi (1 - m \omega_c/\omega)$ 
where $m$ is the integer part of $\omega/\omega_c$. 
The chaos parameter of the map is $K = 4 \epsilon \omega/\omega_c$ 
and the resonance 
separatrix width $\delta v_y = 4 \sqrt{\epsilon \omega_c/\omega}$. The 
energy barrier of the resonance is given by 
$E_r = (\delta v_y)^2 / 2 =8\epsilon \omega_c/\omega$.

In presence of weak dissipation the center of resonance acts as an attractor 
for trajectories inside the resonance. The presence of small angle scattering 
leads to a broadening of the attractor but 
trajectories are still trapped inside. 
If the center is located near $v_y = 0$ particles 
are easily kicked out from the edge,
transmission $T$ drops and $R_{xx}$ increases. On the other hand, 
if the resonance 
width $\delta v_y$ does not touch $v_y = 0$ 
then orbits trapped inside propagate ballistically
with $T \rightarrow 1$ and $R_{xx} \rightarrow 0$. 
The trapping is confirmed in Figs.~1e,f for both models 
at $\omega/\omega_c = 9/4$
with propagating trajectories concentrated inside the resonance,
whereas for $\omega/\omega_c = 2$ in Fig.~1d the region inside the resonance 
does not propagate (propagating orbits concentrate on the unstable separatrix
and their number is much smaller).

\begin{figure}
\begin{center}
\includegraphics[clip=true,width=8.5cm]{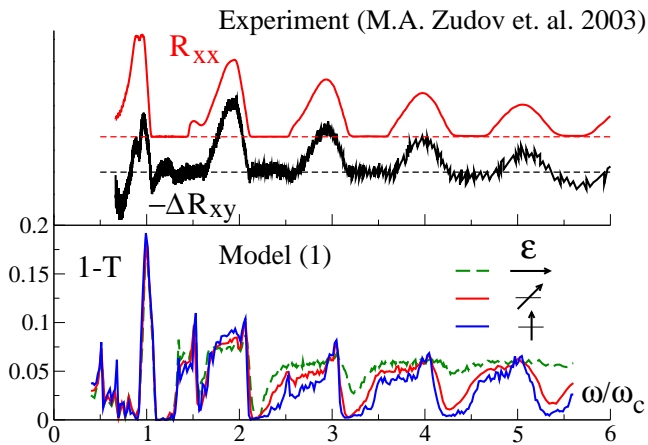}
%\leavevmode
%\epsfxsize=8cm
\vglue -0.3cm
\caption{(Color online) Top panel: dependence of $R_{xx}$ 
and $-\Delta R_{xy}$ (in arbitrary units) on 
$\omega/\omega_c$ from \cite{zudov2003}. 
$\Delta R_{xy}$ is obtained from measured Hall resistance 
by subtracting a linear fit to $R_{xy}$.
Bottom panel: calculated transmission along sample edge 
for three microwave polarizations axis. 
Microwave field is $\epsilon = 0.05$, relaxation 
$\gamma_0 = 10^{-3}$ and noise amplitude $\alpha = 3\times10^{-3}$. 
Transmission without microwaves is $T \simeq 0.95$. 
\label{fig2}}
\end{center}
\end{figure}   

In order to compare our theory with experiment we calculate 
the transmission $T$ for model (\ref{newton}).
An ensemble of $N = 5000$ particles is thrown on the wall at $x = 0$
 with random velocity angle.
They propagate in positive $x$ direction but due to noise some 
trajectories detach from the wall,
we consider that a particle is lost in the bulk when it does not collide 
with the wall for time $20 \pi / \omega_c$. These particles do not 
contribute to transmission 
which is defined as the fraction of particles that reaches $x = 250 v_F/\omega$,
that can be viewed as a distance between contacts.
For $l_e \gg r_c$ the billiard model of a Hall bar \cite{beenakker,lps} gives 
$R_{xx} \propto 1-T$ and a deviation from the classical Hall conductance 
$\Delta R_{xy} = R_{xy} - B / n e \propto -(1-T)$. 
The data in Fig.~2 show calculated $1-T$ and experimental $R_{xx}$ and 
$\Delta R_{xy}$ \cite{zudov2003}.
One can see a good agreement between results of model 
(\ref{newton}) and experimental data.
Both show $R_{xx}$ peaks at integer $j$ and zeros around $j=5/4,9/4,...$.
We also reproduce peaks and dips for ``fractional'' 
ZRS around $j=3/2,1/2$ \cite{zudov2006}.
Our specular wall potential is specially suited for 
the cleaved samples from \cite{zudov2003}
where edges should follow  crystallographic directions but 
peak positions can be shifted for other edge potentials. 
We also note that the possibility to observe ZRS on 
$\Delta R_{xy}$ was discussed in \cite{mani2004}.
Finally our data show weak dependence on polarization axis which supports the 
Chirikov standard map model.

\begin{figure}
\begin{center} % fit with theory.
\includegraphics[clip=true,width=8.5cm]{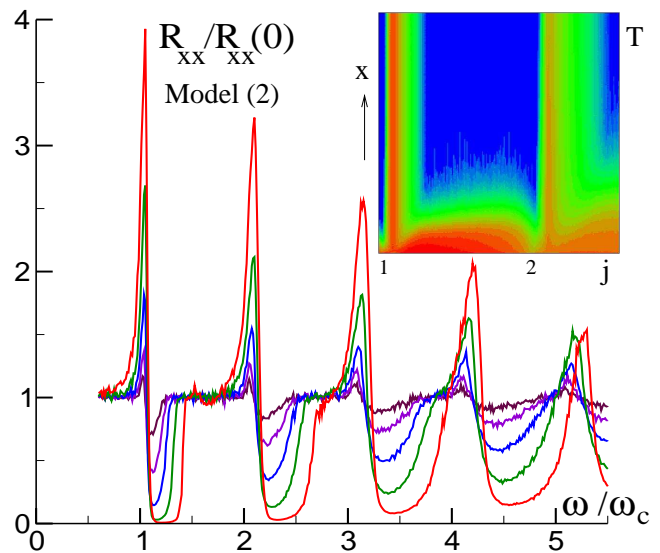}
%\leavevmode
\vglue -0.3cm
\caption{ (Color online) Dependence of rescaled $R_{xx}$
in model (\ref{chirikovmap}) on $\omega/\omega_c$
for microwave field  $\epsilon = 0.00375$, $0.0075$,
$0.015$, $0.03$, $0.06$ 
(curves from top to bottom at $j=\omega/\omega_c=4.5$);
$\gamma_c=0.01$, $\alpha=0.03$. Average is done over 
$10^4$ particles and 5000 map iterations.
Insert shows transmission probability $T$ at
distance $x$ along the edge for $\epsilon=0.02$
(red/gray is for maximum, blue/black for zero, $0<x<10^3v_F/\omega$). 
\label{fig3}}
\end{center}
\end{figure}

Model (\ref{chirikovmap}) is more accessible to analytical analysis and 
numerical simulations. 
In this model a particle is considered lost in the bulk as soon as $v_y < 0$.
The displacement along the edge between collisions is 
$\delta x =2 v_y /\omega_c$
and an effective ``diffusion'' along the edge is defined as
$D_x(\epsilon)=(\Delta x)^2/\Delta t$ where 
$\Delta x$ is a total displacement along the edge during the computation time 
$\Delta t \sim 10^4 /\omega$.
In numerical simulations $D_x$ is averaged over $10^4$ particles homogeneously
distributed in phase space. We then assume that $R_{xx} \propto 1/D_x$
and present the dependence of the dimensionless ratio 
$R_{xx}/R_{xx}(\epsilon =0)$
on $\omega/\omega_c$ in Fig.~3. The computation of transmission $T$ 
(shown in Fig.~3 inset)
gives similar results but is less convenient for numerical analysis. The 
dependence on $j=\omega/\omega_c$ is similar to those shown in Fig.~2.
Both peaks and  dips grow with the increase of microwave field $\epsilon$.

The dependence on $\epsilon$ can be understood from the following arguments. 
Due to noise a typical spread square width in 
velocity angle during the relaxation time 
$1/\gamma_c$ is $D_s = \alpha^2/\gamma_c$. The resonance square width 
is $(\delta v_y)^2 = 16 \epsilon \omega_c/\omega$ and therefore the probability 
to escape from the resonance is 
\begin{align}
W \sim \exp(-(\delta v_y)^2/D_s) \sim \exp(-A \epsilon \omega_c/(D_s \omega))
\label{eqW}
\end{align}
Edge transport is ballistic for exponentially small $W$ and  
$R_{xx}/R_{xx}(0) \sim 1 - T \sim W$. The above estimate gives the numerical 
coefficient $A = 16$ while numerical data presented in Fig.~4 
for model (\ref{chirikovmap})
give $A \approx 12$, and confirm dependence Eq.~(\ref{eqW}) 
on all model parameters.
It holds when edge transport is stabilized by the presence 
of the nonlinear resonance 
which corresponds to regions around $j = 5/4,9/4,...$.
Deviations appear when the parameter 
$K = 4 \epsilon \omega/\omega_c$ approaches 
the chaos border $K \approx 1$ and trapping is weakened by chaos. 
The numerical data for model (\ref{newton}) based on transmission 
computation confirm the scaling dependence
 $\log R_{xx}/R_{xx}(0) \propto -\omega_c \epsilon/\omega$
as shown in Fig.~4. This dependence holds also for other 
models of dissipation in Eqs.~(\ref{newton},\ref{chirikovmap}).
It is consistent with the power dependence 
measured in \cite{mani2002}. A detailed analysis of
 the power dependence may be complicated 
due to heating and out of equilibrium effects at strong power, 
but the global exponential decay of $R_{xx}$ 
with power was confirmed in \cite{mani2004}.

\begin{figure}
\begin{center}
\includegraphics[clip=true,width=8cm]{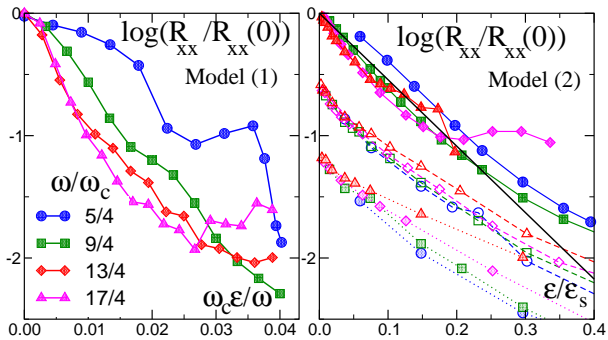}
\vglue -0.3cm
\caption{(Color online) Dependence of rescaled $R_{xx}$
on rescaled microwave field $\epsilon$
for models (\ref{newton}) (left) and
(\ref{chirikovmap}) (right). Left: parameters as in Fig.~2 and $\epsilon$
is varied. 
Right: $\gamma=0.01$, $\alpha=0.02$ (full curves),
$\gamma=0.01$, $\epsilon=0.03$ (dashed curves),
$\epsilon=0.03$, $\alpha=0.02$ (dotted curves), 
the straight line shows theory (\ref{eqW}) with $A=12.5$;
symbols are shifted for clarity and $\epsilon_s=\omega D_s/\omega_c$.
Logarithms are decimal.
\label{fig4}}
\end{center}
\end{figure}

The billiard model used in our studies focuses on dynamics of an electron 
on the Fermi surface 
which corresponds to a zero temperature limit. In order to include 
the effect of temperature $T_e$
one needs to account for the thermal smearing of the electrons around 
the Fermi surface. 
The relaxation rate to the Fermi surface that we introduced 
in our model is also likely 
to depend on temperature. This makes rigorous analysis 
of temperature dependence challenging. 
A simple estimate can be obtained in the frame of 
Arrhenius law with activation energy 
equal to the energy height of the nonlinear resonance 
$E_r = 16 \epsilon \omega_c E_F/\omega$
where $E_F$ is the Fermi energy. This dependence appears as 
an additional damping factor 
in ZRS amplitude in a way similar to temperature 
dependence of Shubnikov-de Hass oscillations leading to
\begin{align}
R_{xx} \propto  \exp(-A \epsilon \omega_c/(D_s \omega)) 
\exp(- 16 \epsilon \omega_c E_F/\omega T_e)
\end{align}
Our prediction on activation energy $E_r$ is in a good agreement 
with experimental data 
and reproduces the proportionality dependence 
on magnetic field observed 
in \cite{mani2002,zudov2003}. For a typical $\epsilon = 0.01$ 
we obtain $E_r \sim 20\;{\rm K}$ at
$j=1$. The proposed mechanism can find applications for microwave induced 
stabilization of ballistic transport 
in magnetically confined quantum wires \cite{portal}.

In summary we have shown that microwave radiation can stabilize 
edge trajectories against 
small angle disorder scattering. For propagating edge channels 
a microwave field creates a nonlinear 
resonance well described by the Chirikov standard map. Dissipative 
processes lead to trapping 
of particle inside the resonance. 
Depending on the position of the resonance center 
in respect to the edge the channeling of particles can be enhanced or weakened  
providing a physical explanation of ZRS dependence on 
the ratio between microwave 
and cyclotron frequencies. In the trapping case transmission along 
the edges is exponentially 
close to unity, naturally leading to an exponential drop in 
$R_{xx}$ with microwave power. 
Our theory also explains the appearance of large energy scale 
in temperature dependence of ZRS. 
A complete theory should also take into account quantum effects since 
about ten Landau levels are typically captured inside the resonance. 
A microscopic 
treatment of dissipation mechanism is also needed for 
further theory development. 

We are grateful to H. Bouchiat, A.A. Bykov and A.S. Pikovsky for fruitful 
discussions. This research is supported in part by 
the  $\;$ ANR  $\;$ PNANO $\;$ project $\;$ NANOTERRA, A.D.C acknowledges DGA for support. 

\vglue -0.3cm

\end{document}